\documentclass[twoside,draft,reqno,12pt]{amsart}
\usepackage[english]{babel}
\usepackage{amsmath}
\usepackage{amssymb}
\usepackage{enumerate}

\usepackage{amsfonts}
\usepackage{graphicx}

\usepackage[ citecolor=black, linkcolor=black,
colorlinks, hypertexnames=false, plainpages=false]{hyperref}

\usepackage{color}

\newtheorem{lemma}     {Lemma}[section]
\newtheorem{thm}   [lemma]{Theorem}
\newtheorem{prop}       [lemma]{Proposition}

\newtheorem{cong1}      [lemma]{Conjecture}
\newtheorem{remark1}    [lemma]{Remark}

\numberwithin{equation}{section}

\newcommand{\und}{\underline}

     \newcommand{\nn}{\nonumber}

\newcommand{\dis}{\displaystyle}

\newcommand{\mmmintone}[1]{{\dis{\int\kern -.38cm
-}}_{\kern-.21cm\substack{#1}}\;\;}
\newcommand{\mmmintwo}[2]{{\dis{\int\kern -.43cm
-}}_{\kern-.21cm\substack{#1}}^{\substack{#2}}\;\;}
\newcommand{\submint}{{\scriptstyle{\int\kern -.66em -}}}
\newcommand{\submintone}[1]{{\scriptstyle{\int\kern -.66em
-}}_{\scriptscriptstyle{\kern-.21em\substack{#1}}}}
\newcommand{\fracmint}{{\textstyle{\int\kern -.88em -}}}
\newcommand{\fracmintone}[1]{{\textstyle{\int\kern -.88em
-}}_{\scriptscriptstyle{\kern-.21em\substack{#1}}}\;}

\newcommand{\eps}{\epsilon}

\newcommand{\la}{\lambda}
\newcommand{\La}{\Lambda}

\newcommand{\E}{\mathbb E}

\newcommand{\nada}[1]{}

\begin{document}
\today

\vskip.5cm
\title
{Non equilibrium stationary state for the SEP with births and deaths}
\begin{abstract}
We consider the symmetric simple exclusion process in the interval $\La_N:=[-N,N]\cap\mathbb Z$ with births and deaths taking place respectively on suitable boundary intervals $I_+$ and $I_-$, as introduced in De Masi et al. (J. Stat. Phys. 2011). We study the stationary measure density profile in the limit $N\to\infty$.

\end{abstract}

\author{A. De Masi}
\address{Anna De Masi,
Dipartimento di Matematica, Universit\`a di L'Aquila \newline
\indent L'Aquila, Italy}
\email{demasi@univaq.it}

\author{E. Presutti}
\address{Errico Presutti,
Dipartimento di Matematica, Universit\`a di Roma Tor Vergata \newline
\indent Roma, 00133, Italy}
\email{Presutti@mat.uniroma2.it}

\author{D. Tsagkarogiannis}
\address{Dimitrios Tsagkarogiannis,
Dipartimento di Matematica, Universit\`a di Roma Tor Vergata \newline
\indent Roma, 00133, Italy}
\email{tsagkaro@mat.uniroma2.it}

\author{M.E. Vares}
\address{ Maria Eulalia Vares,
Instituto de Matem\'atica - Universidade Federal do Rio de Janeiro \newline \indent Av. Athos da Silveira Ramos 149,  21941-909 Rio de Janeiro - RJ - Brasil}
\email{eulalia@cbpf.br}
\maketitle

\section{Introduction}
\label{intro}

This paper is a follow-up of the study initiated in \cite{DPTVjsp}, \cite{DPTVpro}, where current reservoirs in the context of stochastic interacting particle systems have been proposed as a method to investigate stationary non-equilibrium states with steady currents produced
by action at the boundary.

Due to the particular difficulties in implementing this new method, we consider the simplest possible particle system.
The bulk dynamics is the symmetric simple exclusion process (SSEP) in the interval $\La_N=[-N,N]\cap \mathbb Z$ ($N$ a positive integer and $N\to \infty$ eventually), namely  the state space is $\{0,1\}^{\La_N}$ (at most one particle per site): independently each particle tries to jump  at rate $N^2/2$ to each one of its nearest neighbor (n.n.) sites,
the jump then takes place if and only if the chosen site is empty, jumps outside $\La_N$ are suppressed.
To induce a current we send in
particles from the right and take them out from the left, and would like this to happen at rate $Nj/2$,
$j>0$ a  fixed parameter independent of $N$. Due to the restrictions imposed by the configurational space, we have to be more precise when defining this dynamics. For this we fix a parameter $K\ge 1$ (an integer) and two intervals
$I_{\pm}$ of  length $K$  at the boundaries: $I_+\equiv [N-K+1,N]$ and $I_-\equiv [-N,-N+K-1]$.
At rate $Nj/2$, when $I_+$ is not totally occupied, we create a particle at its rightmost empty site; with the same rate, unless $I_-$ is empty, we take out a particle from its leftmost occupied site.  In case $I_+$ is already full, or  $I_-$  empty, the corresponding mechanism aborts.

\smallskip

In \cite{DPTVjsp}, \cite{DPTVpro} we have proved that  at any time $t>0$ propagation of chaos holds and that in the limit $N\to \infty$ the hydrodynamical equation is  the linear heat equation:
		\begin{eqnarray}
	\nn&&\frac {\partial}{\partial t} \rho(r,t)= \frac 12 \frac {\partial^2}{\partial r^2} \rho(r,t), \qquad r\in (-1,1),\, t>0,
	\\&&\rho(r,0)=\rho_0(r),\qquad \rho(\pm 1,t)= u_{\pm}(t),
	\label{1.1}
	\end{eqnarray}
where $\rho_0(\cdot)$ is given but $u_{\pm}(t)$  are solutions of a nonlinear system of two integral equations, see \eqref{dptv2.4} below.
\smallskip

The goal of this paper is to investigate the limiting density profile (as $N\to \infty$) of the (unique) invariant measure of the process. The main result is Theorem \ref{thm2.2}, which shows that this rescaled limiting profile coincides with the unique stationary solution of \eqref{1.1}. In particular, taking into account the validity of the Fourier law, proven as Theorem 2 in \cite{DPTVjsp}, we see that the effective current in the stationary regime is strictly smaller than its desired maximum value which is $\min\{j/2,1/4\}$, but this value is indeed approached by letting $K\to \infty$.
\vskip1cm

\section{Model and main results}

Particle configurations are elements $\eta$ of $\{0,1\}^{\La_N}$, $\eta(x)=0,1$ being the occupation number at $x\in\La_N$. We consider the Markov process on $\{0,1\}^{\La_N}$ defined via the  generator
		$$
L_N:=N^{2}\Big(L_0+\frac 1N L_b\Big),
		$$
where $L_b= L_{b,+}+ L_{b,-}$ and
      \begin{eqnarray}
        \nn
&&
L_0 f(\eta):=\frac 12\sum_{x=-N}^{N-1} [f(\eta^{(x,x+1)})-f(\eta)],
\\ \label{1}
\\&& L_{b,\pm} f(\eta):= \frac{j}{2}\sum_{x\in I_\pm}D_{\pm}\eta(x) [f(\eta^{(x)})-f(\eta)\Big],
\nn
     \end{eqnarray}
$\eta^{(x)}$ being the configuration obtained from $\eta$ by changing the occupation number at $x$, $\eta^{(x,x+1)}$ by exchanging the occupation numbers at $x,x+1$; for any $u:\La_N\to [0,1]$
           \begin{eqnarray}
&& D_+u (x)= [1-u(x)]u(x+1)u(x+2)\dots u(N), \quad  x\in I_+
       \nn\\
&& D_-u (x)=  u(x)[1-u(x-1)][1-u(x-2)]\dots[1-u(-N)], \quad  x\in I_-.
       \label{6}
            \end{eqnarray}

\vskip 0.5cm

Given  $\rho_0\in C([-1,1],[0,1])$, let $\nu^{(N)}$ be the product probability measure on $\{0,1\}^{\La_N}$ such that $\dis{\nu^{(N)}(\eta(x))=\rho_0(N^{-1} x)}$ for all $x\in\La_N$. Let $\mathbb{P}_{\nu^{(N)}}$ denote the law of the process with initial distribution $\nu^{(N)}$ and $\E_{\nu^{(N)}}$ the corresponding expectation.\footnote{Omitting the initial profile to avoid too heavy notation.}

\vskip.5cm
The following theorem has been proven (in a stronger form) in \cite{DPTVjsp}, \cite{DPTVpro}. The statement below contains all what is needed in the present paper. In the following, for $n$ a positive integer we write $\Lambda_N^{n,\neq}$ for the set of all sequences  $(x_1,...,x_n)$ in $\La_N^n$ such that   $x_i\ne x_j$ whenever $i\neq j$.
\vskip.5cm
  \begin{thm}
                \label{thm2.1}
There exists $\tau>0$ so that for any $\rho_0$ as above and any $n\ge 1$,
     \begin{equation}
             \label{2.0}
             \lim_{N\to \infty} \Big|\E_{\nu^{(N)}}\big(\prod_{i=1}^n \eta(x_i,t)\big)-\prod_{i=1}^n\E_{\nu^{(N)}}\big( \eta(x_i,t)\big)\Big|=0,\qquad \text{for any  } t\le \tau\log N.
        \end{equation}
Furthermore
             \begin{equation}
             \label{dptv2.3}
\lim_{N\to \infty} \sup_{x\in \La_N}\sup_{ t \le \tau\log N} \big|\E_{\nu^{(N)}}\big( \eta(x,t) )-\rho( N^{-1}x,t)\big|=0,
             \end{equation}
where the function $\rho(r,t)$  solves the heat equation $\frac{\partial \rho}{\partial t}=\frac 12 \frac{\partial^2 \rho}{\partial r^2}$, $r\in (-1,1), t>0$  with initial datum $\rho_0$ and boundary conditions $\rho(\pm 1,t)=u_\pm(t)$, the pair $(u_+(t),u_-(t))$ being the unique solution of the non linear system
	           \begin{eqnarray}
            \nn
&&
\hskip-.3cm
u_\pm(t)= \int_{[-1,1]} P_t(\pm 1,r) \rho_0(r) dr+ \frac j2\int_0^t \Big\{ P_{s}(\pm 1,1)\left(1-u_+(t-s)^K\right)
\\&&\hskip3.8cm
- P_{s}(\pm 1,-1)\left(1- (1-u_-(t-s))^K\right)\Big\}ds,
			 \label{dptv2.4}
      \end{eqnarray}
where $P_t(r,r')$  is the density kernel of the semigroup (also denoted as ${P_t}$) with generator $\Delta/2$, $\Delta$ the laplacian in $[-1,1]$ with reflecting, Neumann, boundary conditions.

The function $\rho(r,t)$ satisfies
			\begin{equation}
            \label{4.2}
\frac{\partial \rho(r,t)}{\partial r}|_{r=1}=j (1-u_+(t)^K),\quad \frac{\partial \rho(r,t)}{\partial r}|_{r=-1}=j (1-(1-u_-(t))^K).
             \end{equation}

	\end{thm}
\vskip.5cm
\noindent {\bf Remark.} The following is the integral form of the macroscopic equation:
	\begin{eqnarray}
             \label{dptv2.4bis}
&& \rho(r,t)= \int_{[-1,1]} P_t(r,r') \rho(r',0) dr' + \frac j2\int_0^t \Big\{ P_{s}(r,1)\left(1-\rho(1,t-s)^K\right)\nn
\\&&\hskip3cm
- P_{s}(r,-1)\left(1- (1-\rho(-1,t-s))^K\right)\Big\}ds.
      \end{eqnarray}
It will be convenient to recall the expression for the density kernel $P_t(r,r^\prime)$ in terms of the Gaussian kernel
\begin{equation}
    \label{N4.3}
 G_t (r,r')  = \frac{e^{-(r-r')^2/(2t)}}{\sqrt {2\pi t}},\quad r,r' \in \mathbb{R},
    \end{equation}
    as
  \begin{eqnarray}
    \label{N4.10}
           P_t (r,r') &=& \sum_{r'':\psi(r'')=r'} G_t(r,r'') \quad \text{for} \; r'\neq \pm 1\nn\\
           P_t(r,\pm 1) &=& \sum_{r'':\psi(r'')=\pm 1} 2G_t(r,r''),
 \end{eqnarray}
where $\psi:\mathbb R\to [-1,1]$ denotes the usual reflection map:
$\psi (x)= x $ for $x \in [-1,1]$, $\psi(x)=2-x $ for $x \in [1,3]$,  $\psi$ extended to the whole line
as periodic of period 4.
\vskip .5cm
\noindent {\bf Notation.} $P_t g(r)=\int P_t(r,r^\prime) g(r^\prime) dr^\prime$, for $g$ a bounded continuous function, $t>0$.

\vskip1cm

\noindent
The main result of this paper is about the density profile of the unique invariant measure $\mu_N$.

                \begin{thm}
                \label{thm2.2}
For any integer $k\ge 1$ we have
	\begin{equation}
	\label{2.1-a}
\lim_{N\to\infty} \max_{(x_1,..,x_k)\in \La_N^{k,\ne}}\Big|\mu_N\big(\eta(x_1)\cdots \eta(x_k)\big)-\rho^*(x_1/N)\cdots \rho^*(x_k/N)\Big| = 0
\end{equation}
where $\rho^*(r)$ is the unique stationary solution of the macroscopic equation.
Namely $\rho^*(r)=J\,r +\frac 12$,
	\begin{equation}
	\label{2.1}
J=j(1-\alpha^K),\quad \text{ with $\alpha$ the solution of}\quad \alpha(1+j\alpha^{K-1})=j+\frac 12.
\end{equation}

	\end{thm}
%
%

\vskip.5cm

By Theorem \ref{thm2.2} it   follows that $\mu_N$ concentrates on a $L^1$-neighborhood  of the
limit profile $\rho^*$: let $r\in (0,1)$ and
	\begin{equation*}
\rho^{(\ell)}(r;\eta) = \frac1{2\ell+1} \sum_{x\in \La_N: |x-rN| \le \ell} \eta(x)
\end{equation*}
Then for any $a\in (0,1)$
	\begin{equation*}
\lim_{N\to \infty} \mu_N \Big(  \int_{-1}^1 | \rho^{(N^a)}(r;\eta) -\rho^*(r)|dr\Big)=0
\end{equation*}

Theorem \ref{thm2.2} will follow from  $\bullet$\; uniformly on the initial datum $\rho_0$ the solution $\rho(r,t|\rho_0)$ of the macroscopic equation \eqref{dptv2.4bis} converges in sup norm to $\rho^*$ exponentially fast, see Theorem \ref{thm3.3} below;  $\bullet$\;
for any integer
$k\ge 1$,
	\begin{equation}
	\label{2.1-b}
\lim_{t\to \infty}\lim_{N\to\infty}\max_{\eta \in \{0,1\}^{\La_N}}\max_{(x_1,..,x_k)\in \La_N^{k,\ne}}\Big|\mathbb E_\eta \Big(\prod_{i=1}^k\eta(x_i,t)\Big)-\prod_{i=1}^k\rho^*(x_i/N)\Big| = 0
\end{equation}

We are also working on an extension of the theorem where we prove exponential convergence in time to $\mu_N$  uniformly in $N$.


%

\vskip1cm
\section{Monotonicity properties}
\vskip.5cm
We consider the space $\{0,1\}^{\La_N}$ endowed with the usual partial order, namely we say that $\eta\le \xi$ iff $\eta(x)\le \xi(x)$ for all $x\in \La_N$.
The following proposition is an immediate consequence of general facts on attractive systems, see e.g. \cite{li} (chs. II and III).

           \begin{prop}
                \label{thm3.1}
Let $\eta_0$ and $\xi_0$ be two particle configurations such that $\eta_0\le \xi_0$, and let $\mathbb{P}_{\eta_0}$, respectively
$\mathbb{P}_{\xi_0}$, be the law of the process starting from $\eta_0$, respectively $\xi_0$. Then there is a coupling $\mathbb{Q}$ of $\mathbb{P}_{\eta_0}$ and $\mathbb{P}_{\xi_0}$ (i.e. $\mathbb{Q}$ is a measure on the product space, with $\mathbb{P}_{\eta_0}$ as its
first marginal, and $\mathbb{P}_{\xi_0}$ as the second one) such that
	\begin{equation}
	\label{3.1}
\mathbb{Q}\{(\eta,\xi)\colon \eta_t \,\le \,\xi_t\,, \forall t\}=1
\end{equation}
	\end{prop}

\noindent {\bf Proof}. Being well known that the process corresponding to $L_0$ is attractive, it suffices to observe that the flip rates $c(x,\eta):=D_\pm \eta (x)$ in $I_\pm$ are attractive in the sense that if $\eta(x)=\xi(x)=0$ and $\eta \le \xi$ then $c(x,\eta) \le c(x,\xi)$, while if $\eta(x)=\xi(x)=1$ and $\eta \le \xi$ then $c(x,\xi) \le c(x,\eta)$. \qed

\vskip.5cm

The analogous monotonicity property holds for the macroscopic equation.
Instead of a direct proof
we   derive the result as a   consequence of the monotonicity of the particle system and that it converges
to the macroscopic equation.

\vskip.5cm

 \begin{thm}
                \label{thm3.2}
Let $ \rho_0, \tilde\rho_0$ be bounded measurable functions from $[-1,1]$ to $[0,1]$ such that $\rho_0(r)\le \tilde\rho_0(r)$ for all $r\in[-1,1]$, and let $\rho(r,t)$, respectively $\tilde\rho(r,t)$, be the corresponding solution of \eqref{dptv2.4bis} with initial datum $\rho_0$, respectively $\tilde\rho_0$. Then $\rho(r,t)\le \tilde\rho(r,t)$ for all $r\in[-1,1]$ and $t\ge 0$.
	\end{thm}

\noindent {\bf Proof}.  Let $\nu^{(N)}$ and $\tilde\nu^{(N)}$ be the product probability measures on $\{0,1\}^{\La_N}$ such that $\nu^{(N)}(\eta(x))=\rho_0(N^{-1} x)$ and $\tilde\nu^{(N)}(\eta(x))=\tilde\rho_0(N^{-1} x)$ for all $x\in\La_N$.
It is well known that a coupling $\la^{(N)}$ of  $\nu^{(N)}$ and $\tilde\nu^{(N)}$ such that $\la^{(N)}\{(\eta,\tilde \eta)\colon \eta\le \tilde\eta\}=1$ exists. Using Proposition \ref{thm3.1} and the notation of Theorem \ref{thm2.1} we have
		\begin{equation}
            		 \label{3.2-a}
\mathbb{E}_{\nu^{(N)}}(\eta(x,t))\le \mathbb{E}_{\tilde\nu^{(N)}}(\eta(x,t)),\qquad \forall x\in \La_N,\quad \forall t\ge 0.
            		 \end{equation}
From \eqref{dptv2.3} we then have that for all  $t\ge 0$ and for all $r\in[-1,1]$, (below $[\cdot]$ denotes the integer part)
		 \begin{equation}
            		 \label{3.2}
\rho(r,t)=\lim_{N\to\infty}\mathbb{E}_{\nu^{(N)}}\big(\eta([Nr],t)\big)\le \lim_{N\to\infty}\mathbb{E}_{\tilde\nu^{(N)}}\big(\eta([Nr],t)\big)=\tilde\rho(r,t).
            		 \end{equation}
\qed

\vskip1cm
\section{The macroscopic profile}
\vskip.5cm
\nopagebreak
We  first prove that the function $\rho^*$ in the statement of Theorem \ref{thm2.2} is a stationary solution  to the Dirichlet problem \eqref{1.1} with boundary condition \eqref{dptv2.4} or, equivalently, of the integral equation \eqref{dptv2.4bis}. In fact by requiring that a stationary solution is a linear function we get, due to  \eqref{4.2}, that the values of this function at $\pm 1$, denoted with $u_\pm$, must satisfy
	\begin{equation*}
j (1-u_+^K) =j (1-(1-u_-)^K).
					\end{equation*}
This implies
	\begin{equation*}
 u_+=(1-u_-),\qquad \text{and }\quad \frac {2u_+-1}2=j(1-u_+^K),\quad u_+=\frac 12 +j(1-u_+^K).
             \end{equation*}
 Solving we get
 	\begin{equation*}
u_+(1+ju_+^{K-1}) =j +\frac 12
					\end{equation*}
in agreement with \eqref{2.1}.

On the other hand, since $\frac{\partial}{\partial t}P_t(r,r')=\frac12\frac{\partial^2}{\partial(r')^2}P_t(r,r')$ and
it satisfies Neumann boundary conditions at $\pm 1$ we easily see that
\begin{equation*}
\frac{d}{dt} \int_{[-1,1]} P_t(r,r')r'dr'=\frac 12\left(P_t(r,-1)-P_t(r,1)\right).
\end{equation*}
Recalling (from \eqref{2.1}) that $J=j(1-(\rho^*(1))^K)=j(1-(1-\rho^*(-1))^K)$
we see at once that $\rho^*$ satisfies \eqref{dptv2.4bis}, which in this case can be written as:
\begin{equation}
\rho^*(r)= P_t\rho^*(r)+\frac j2 (1-(\rho^*(1))^K)\int_0^t \Big\{ P_{s}(r,1)
- P_{s}(r,-1)\Big\}ds,
		  \label{4.1}
      \end{equation}
 for all $t\ge 0$.
\vskip.5cm

We now prove that any solution to the Dirichlet problem converges exponentially fast to $\rho^*$ as $t\to \infty$. In particular, one has uniqueness of the stationary solution.

\vskip.5cm
 	\begin{thm}
                \label{thm3.3}
There exist positive constants $c,c^\prime$ so that for any function $\rho_0$ $\in L^\infty([-1,1],[0,1])$ the solution  $\rho(r,t|\rho_0)$ of the
macroscopic equation \eqref{dptv2.4bis} with initial datum $\rho(r,0)=\rho_0(r)$ satisfies
	\begin{equation}
	\label{t4.1}
\sup_{r\in[-1,1]}|\rho(r,t|\rho_0)-\rho^*(r)|\le c' e^{-ct}.
	\end{equation}
	\end{thm}

\noindent {\bf Proof}.
Let $\bar \rho(r,t)$ denote the solution with initial datum $\rho\equiv 1$, and $\und \rho(r,t)$ that
corresponding to initial datum $\rho\equiv 0$.  From Theorem \ref{thm3.2} we know that
$\und \rho(r,t)\le \rho(r,t|\rho_0)\le \bar \rho (r,t)$, for any initial $\rho_0$. Hence, calling
\begin{equation*}
w(r,t):=\bar\rho(r,t)-\und\rho(r,t) \ge 0,\qquad w(t)=\sup_{r\in[-1,1]}w(r,t)
\end{equation*}
it suffices to show that $w(t)\le c^\prime e^{-ct}$ for suitable positive constants $c,c^\prime$ and all $t>0$.

In the proof below $c, \bar c, \tilde c$ will denote suitable positive constants (that might depend on the model parameter $j$)
whose value may change from line to line.
Let
\begin{equation*}
\bar u_\pm(t):=\bar\rho(\pm 1,t),\quad \und u_\pm(t):=\und\rho(\pm 1,t), \qquad w_\pm(t):=\bar u_\pm(t)-\und u_\pm (t)\ge 0.
\end{equation*}


>From \eqref{dptv2.4bis} we see that for all $r\in[-1,1]$, and all $t\ge t_0\ge 0$,
		  \begin{equation}
            \label{4.3a}
w(r,t)=(P_{t-t_0}w(\cdot,t_0))(r)- \frac j2\int_{t_0}^t  f(r,s,t-s) ds,
	\end{equation}
where
 \begin{eqnarray}
	 	\nn
&&\hskip-.8cm
f(r, s,t-s):= P_{s}(r,1)\left\{\bar u_+(t-s)^K-\und u_+(t-s)^K\right\}
\\&&\hskip1cm +P_{s}( r,-1)\left\{(1-\und u_-(t-s))^K-(1-\bar u_-(t-s))^K\right\}.
	\label{4.0a}
      \end{eqnarray}
 Interchanging particles and holes, one can couple at once the evolutions starting from the configurations $\bar\eta=\und 1$ ({\it all occupied sites})  and $\und \eta =\und 0$ ({\it all empty sites}) so that $\bar \eta(x,t)=1-\und \eta(-x,t)$. Therefore, by the
   same argument as in the proof of Theorem \ref{thm3.2} one has $\bar \rho(r,t)=1-\und\rho(-r,t)$ for all $r$ and all $t$. In particular $w(-r,t)=w(r,t)$,  $\bar u_{\pm}(t)=1-\und u_{\mp}(t)$ and $w_+(t)=w_-(t)$ for all $t$. (Still from Theorem \ref{thm3.2} we see that
   $w(r,t)$ and so also $w(t)$ decrease in $t$.) Of course $w(r,0)=1$ for all $r$.

 In particular, we may rewrite \eqref{4.3a}  with $t_0=0$ as
 \begin{equation}
w(r,t)=1- \frac j2\int_0^{t}  [P_{s}(r,1)+P_s(r,-1)]w(1,t-s)h(t-s)ds\label{4.15}
		\end{equation}
where
	 \begin{equation}
	 	h(t-s):=\sum_{\ell=0}^{K-1}\bar u_+(t-s)^{K-1-\ell}\und u_+(t-s)^{\ell}
	\label{4.16}
      \end{equation}
      and where we have used that for any integer $K\ge 1$,
		  \begin{equation}
            \label{4.3a.1}
a^K-b^K=(a-b)\sum_{\ell=0}^{K-1} b^\ell a^{K-1-\ell},\qquad a \ge b \ge 0.
	\end{equation}
Also, from \eqref{4.16} and the monotonicity properties we see that
\begin{equation}
\label{ck}
b:={\rho^*(1)}^{K-1}\le h(t) \le b+K-1=: c_K.
\end{equation}
     The proof will use local times. To this end we introduce
the kernel operators $K^{(\eps)}_{s}$, $\eps>0$:
  \begin{equation*}
K^{(\eps)}_{s} f(r)=\frac 1{\eps}\int_{[-1,-1+\eps]\cup[1-\eps,1]} P_s(r,r')f(r')dr', \quad f \in C([-1,1],\mathbb R).
  \end{equation*}
In particular $K^{(\eps)}_{s} f(r)=K^{(\eps)}_{s} f(-r)$ for all $r \in [-1,1]$. Let  $w^{(\eps)}$ be the solution to the following
integral equation:
  \begin{eqnarray}
	 	\nn	
&&\hskip-1cm
w^{(\eps)}(r,t)=1
- \frac j2\int_0^{t} (K^{(\eps)}_{s} w^{(\eps)}(\cdot,t-s))(r)h(t-s)ds.
	\label{4.15b}
		\end{eqnarray}
We shall next prove that for all $T>0$,
	 \begin{equation}
            \label{4.16'}
\lim_{\eps\to 0}\,\sup_{r\in [-1,1]}\,\sup_{0\le t\le T}\big|w(r,t)- w^{(\eps)}(r,t)\big|=0.
	\end{equation}
Calling
	 \begin{equation}
            \label{4.16a}
\psi(r,t)=\big|w(r,t)- w^{(\eps)}(r,t)\big|,\qquad  \Psi(t)=\sup_{r\in [-1,1]}\psi(r,t)	
    \end{equation}
and using \eqref{ck}, we can write
	  \begin{eqnarray}
\nn
&&\Big|\int_0^{t} \left\{(K^{(\eps)}_{s} w^{(\eps)}(\cdot,t-s))(r)- \{P_{s}(r,1)+P_s(r,-1)\}w(1,t-s)\right\}h(t-s)ds
\Big|\le c\eps\\
&&\nn +c_K
\int_\eps^{t}\left\{\frac 1{\eps}\int_{1-\eps}^1\big|P_s(r,y)-P_s(r, 1)+P_s(-r,y)-P_s(-r,1)\big|dy\right\} \big|w^{(\eps)}(y,t-s)\big|ds \\
&& +c_K\int_\eps^{t} \{P_{ s}(r, 1)+P_s(-r,1)\} \Psi(t-s) ds.
	\label{4.17}
		\end{eqnarray}
Using that for all $y\in[1-\eps,1], \, r\in [-1,1]$
	 \begin{equation}
	\label{4.18}
|P_s(r,y)-P_s(r, 1)|\le c \frac {1-y}{\sqrt {s^3}}, \qquad \forall s\in [\eps,t]
\end{equation}
we see that the second term on the r.h.s. of \eqref{4.17} is bounded above by
\begin{equation*}
\tilde c \int_\eps^{t} \frac 1{\sqrt {s^3}}ds\frac 1{\eps}\int_{1-\eps}^1(1-y)dy\le c'  \sqrt\eps
\end{equation*}
for suitable constants $\tilde c, c'$. We then easily get
 	\begin{equation}
\psi(r,t)\le c_1 \sqrt\eps+c_2 \int_{0}^t  \Psi(s)ds
		\label{4.20}
	\end{equation}
for suitable constants  $c_1,c_2$. By the Gronwall inequality we conclude \eqref{4.16'}.
\vskip.5cm

We now estimate $w^{(\eps)}$. Let $\{B_t\}$ be a standard Brownian motion with reflecting b.c. at $\pm 1$,
with $\mathbb P_r$  denoting its law when $B_0=r$ (and corresponding expectations denoted by $\mathbb E_r$). Then
		\begin{equation}
w^{(\eps)}(r,t)= \mathbb E_r\Big( e^{-\int_0^{t} \varphi_\eps(B_s,t-s)ds} w^{(\eps)}(B_{t},0)\Big)
		\label{4.21}
	\end{equation}
where
		\begin{eqnarray}
\varphi_\eps(B,t-s)= \phi_\eps(B)h (t-s),\quad  \phi_\eps(r)=\frac {j}{2\eps}\mathbf 1_{[1-\eps,1]}(|r|), \; r\in [-1,1].
		\label{4.22}
	\end{eqnarray}
By \eqref{ck}
		\begin{equation}
w^{(\eps)}(r,t)  \le \mathbb E_r\Big( e^{-b \int_0^{t} \phi_\eps(B_s)ds} \Big).
		\label{4.21.1}
	\end{equation}

For  $0<\bar t<t$  we write
		\begin{equation}
w^{(\eps)}(r,t) \le \mathbb E_r\Bigg( e^{-b\int_{0}^{t-{\bar t}} \phi_\eps(B_s)ds}\,
\mathbb E_{B_{t-\bar t}}\Big( e^{-b\int_{t-\bar t}^t \phi_\eps(B_s)ds}\Big)\Bigg)
		\label{4.23}
	\end{equation}
We shall prove below that taking $\bar t$ sufficiently small, we can take $\alpha<1$ so that for all $\eps>0$
	\begin{equation}
\sup_{r\in[-1,1]}
\mathbb E_{r}\Big( e^{-b\int_{0}^{\bar t} \phi_\eps(B_s)ds} \Big)\le 1-\alpha
		\label{4.24}
	\end{equation}
From \eqref{4.24} and \eqref{4.23}  we then get
	\begin{equation}
|w^{(\eps)}(r,t)|\le (1-\alpha)^{[t/\bar t]}
		\label{4.25}
	\end{equation}
($[a]$ the integer part of $a$) which then concludes the proof of the theorem.

\vskip.5cm
{\bf Proof of \eqref{4.24}}. Let $T=\inf\{t \ge 0: |B_t|=1\}$.
We then have
			\begin{eqnarray}
			\label{4.26}
 \mathbb E_{r}\Big( e^{-b\int_{0}^{\bar t} \phi_\eps(B_s)ds} \Big)&\le& \mathbb E_{r}\Big( \mathbf 1_{\{T\le \bar t/2\}}e^{-b\int_{T}^{\bar t} \phi_\eps(B_s)ds} \Big)+\mathbb P_r(T> {\bar t}/2)
    \end{eqnarray}
and write
        \begin{eqnarray}
\mathbb E_{r}\Big( \mathbf 1_{\{T\le {\bar t}/2\}}e^{-b\int_{T}^{\bar t} \phi_\eps(B_s)ds} \Big) &\le& \mathbb E_{r}\Big(\mathbf 1_{\{T \le {\bar t}/2\}}\mathbb E_{B_T}\big( e^{-b\int_0^{{\bar t}/2} \phi_\eps(B_s)ds} \big)\Big) \nn
\\ &\le& \mathbb{P}_r(T \le {\bar t}/2)\;
\mathbb E_{1}\big( e^{-b\int_0^{{\bar t}/2} \phi_\eps(B_s)ds} \big)\Big)
\label{a4.26b}
	\end{eqnarray}
where we also used that $\mathbb E_{1}\big( e^{-b\int_0^{{\bar t}/2} \phi_\eps(B_s)ds} \big)=\mathbb E_{-1}\big( e^{-b\int_0^{{\bar t}/2} \phi_\eps(B_s)ds} \big)$ by symmetry.

 By Taylor expansion
	\begin{equation}
\mathbb E_{1}\big( e^{-b\int_0^{{\bar t}/2} \phi_\eps(B_s) ds}\big)
\le 1-b\mathbb E_{1}\Big(\int_0^{{\bar t}/2} \phi_\eps (B_s)ds \Big)
+\xi_2
		\label{4.27}
	\end{equation}
	where
					\begin{equation}	
\xi_2=(\frac{jb}{2\eps})^2\int_0^{{\bar t}/2} dt_1\int_0^{t_1}dt_2
\int_{|y_1|\in [1-\eps,1], |y_2|\in [1-\eps,1]} P_{t_1}(1,y_1)P_{t_2}(y_1,y_2) dy_1dy_2
		\label{4.28}
	\end{equation}
But, from \eqref{N4.3}--\eqref{N4.10} we see that
		\begin{equation}
		\label{4.29}	
\sup_{x,y\in [-1,1]}P_s(x,y)\le c \frac 1{\sqrt s}
	\end{equation}
so that for $\bar t$ small enough we get
	\begin{equation}
		\label{4.30}
 \xi_2\le \bar c {\bar t}/2
 		\end{equation}
for suitable constant $\bar c$.
Using again \eqref{N4.3}--\eqref{N4.10}, we see at once that a positive constant $c$ can be taken so that for all $\bar t$ small,
and all $\eps>0$
	\begin{equation}
		\label{4.31}
b\mathbb E_{1}\Big(\int_0^{{\bar t}/2} \phi_\eps(B_s)ds \Big)\ge
	c\sqrt {\bar t/2}.
 		\end{equation}
From  \eqref{4.27}, \eqref{4.30} and \eqref{4.31} we then get
for $\bar t$ small (with possibly different constant $c$),
	\begin{equation}
			\label{4.32}
\mathbb E_{1}\big( e^{-b\int_0^{{\bar t}/2} \phi_\eps(B_s)ds} \big)\le  1-c\sqrt{{\bar t}/2}.
	\end{equation}
By \eqref{4.26} and \eqref{a4.26b} we then have
    \begin{equation}
			\label{4.41}
 \mathbb E_{r}\Big( e^{-b\int_{0}^{\bar t} \phi_\eps(B_s)ds} \Big)\le \big[1-c\sqrt{{\bar t} /2}\big]\mathbb{P}_r(T\le {\bar t}/2)+\mathbb{P}_r(T>{\bar t}/2)\le 1-\alpha
    \end{equation}
 with
    \begin{equation}
     \alpha=\inf_{r\in [-1,1]} \mathbb{P}_r(T\le {\bar t}/2)c\sqrt{{\bar t} /2}.
    \end{equation}

	\vskip1cm
\section{Proof of Theorem \ref{thm2.2} }
\vskip.5cm
\nopagebreak
The proof is a direct consequence of the following three facts.  (i) For any $t>0$ and any integer $k\ge 1$
	\begin{equation}
		\label{5.0.1}
	\lim_{N\to \infty} \max_{ \eta \in \{0,1\}^{\La_N}}\max_{(x_1,..,x_k)\in \La_N^{k,\ne}}
\Big| \mathbb E_{\eta}\big(\prod_{i=1}^k \eta(x_i,t)\big)-\prod_{i=1}^k \mathbb E_{\eta}\big(\eta(x_i,t)\big)\Big|=0
		\end{equation}
(ii) For any $t>0$
	\begin{equation}
	\label{5.0.2}
	\lim_{N\to \infty}
\max_{ \eta \in \{0,1\}^{\La_N}}\max_{x\in \La_N}
\Big| \mathbb E_{\eta}\big( \eta(x,t)\big)-\rho(x/N,t| \eta)\Big|=0
		\end{equation}
(iii)
	\begin{equation}
	\label{5.0.3}
	\lim_{t\to \infty}
\sup_{ \rho_0\in L^\infty([-1,1];[0,1])}
\| \rho(\cdot,t| \rho_0)-\rho^*(\cdot)\|_\infty=0
		\end{equation}
(i) and (ii) are proved in \cite{DPTVpro}--\cite{DPTVjsp}, \eqref{5.0.3} is proved in Theorem \ref{thm3.3}.

\vskip.5cm

{\bf Acknowledgments.}

The work is partially supported by the PRIN project, n.2009TA2595. M.E.V. thanks the Universities of Rome and L'Aquila for kind hospitality. M.E.V. is
partially supported by CNPq grant 302796/2002-9.  The research of D.T. has been partially supported by a
Marie Curie Intra European Fellowship within the 7th European
Community Framework Program.

\vskip1cm

\bibliographystyle{amsalpha}

\end{document}